 \documentclass[letterpaper]{aa} 
 \usepackage{graphicx}
 \usepackage{txfonts}
 \usepackage{natbib}
 \bibpunct{(}{)}{;}{a}{}{,} 
 \def\msun{{\it M}$_{\sun}$}
 
\begin{document}
 %
 
    \title{Reaction rate uncertainties and $^{26}$Al in AGB silicon carbide stardust}
 
    \author{M. A. van Raai
           \inst{1}
           \and
           M. Lugaro
 	  \inst{1,2}
 	  \and
           A. I. Karakas
 	  \inst{3}
 	  \and
           C. Iliadis
 	  \inst{4}
           }
 
    \institute{Sterrekundig Instituut, University of Utrecht,
               Postbus 80000 3508 TA Utrecht, The Netherlands\\
               \email{m.a.vanraai@phys.uu.nl, m.lugaro@phys.uu.nl}
          \and
              Centre for Stellar and Planetary Astrophysics, School of Mathematical Sciences,
              Monash University, Victoria 3800, Australia
          \and
 	     Research School of Astronomy and Astrophysics,
 	     Mt. Stromlo Observatory, Cotter Rd., Weston,
 	     ACT 2611, Australia\\
              \email{akarakas@mso.anu.edu.au}
          \and
 	     Department of Physics and Astronomy, University of North Carolina, Chapel
              Hill, NC 27599-3255, USA; Triangle Universities Nuclear Laboratory, P. O.
              Box 90308, Durham, NC 27708-0308, USA\\
               \email{iliadis@unc.edu}
              }
 
    \date{Received / Accepted }
 
    \authorrunning{van Raai et al.}
    \titlerunning{$^{26}$Al in AGB silicon carbide stardust}

 
    \abstract
           { Stardust is a class of presolar grains each of which
 	  presents an ideally
 	  uncontaminated stellar sample.
 	  Mainstream silicon carbide (SiC) stardust
 	  formed in the extended envelopes of carbon-rich asymptotic giant branch
           (AGB) stars and incorporated the radioactive nucleus $^{26}$Al as
           a trace element.}
 	  {The aim of this paper is to analyse in detail the effect of nuclear
 	  uncertainties, in particular the large uncertainties of up to four
 	  orders of
 	  magnitude related to the
 	   $^{26}$Al$_g$($p,\gamma$)$^{27}$Si reaction rate, on the production of
 	  $^{26}$Al in
 	  AGB stars and compare model predictions to data obtained from
 	  laboratory analysis of
 	  SiC stardust grains. Stellar uncertainties are also briefly discussed.}
 	  {We use a detailed nucleosynthesis postprocessing code to
 	  calculate the
 	  $^{26}$Al/$^{27}$Al ratios at the surface of AGB stars of
           different masses ({\it M} = 1.75, 3, and 5 {\it M}$_{\sun}$) and
 	  metallicities ({\it Z} = 0.02, 0.012, and 0.008).}
 	  {For the lower limit and recommended value of the
 	   $^{26}$Al$_g$($p,\gamma$)$^{27}$Si reaction rate, the predicted
 	  $^{26}$Al/$^{27}$Al ratios replicate the upper values
 	  of the range of the $^{26}$Al/$^{27}$Al
           ratios measured in SiC grains.
 	  For the upper limit of the  $^{26}$Al$_g$($p,\gamma$)$^{27}$Si
 	  reaction rate, instead, the predicted
 	  $^{26}$Al/$^{27}$Al ratios are $\approx$ 100 times
 	  lower and lie below the range
 	  observed in SiC grains.
 	  When considering models of different masses and
 	  metallicities, the spread of more
 	  than an order of magnitude in the $^{26}$Al/$^{27}$Al
 	  ratios measured in stellar SiC grains is not reproduced.}
 	  { We propose two scenarios to explain the spread of the
 	  $^{26}$Al/$^{27}$Al ratios observed in mainstream SiC, depending
 	  on the choice of the $^{26}$Al$_g$$+p$ reaction rate. One involves
 	  different times of stardust formation, the other involves
 	  extra-mixing processes.
 	  Stronger conclusions on the
 	  interpretation of the Al composition of AGB stardust will be possible after
 	  more information is
 	  available from future nuclear experiments on the 
	  $^{26}$Al$_g$$+p$ reaction.}
 	
 
    \titlerunning{$^{26}$Al production in AGB stars}
 
    \maketitle
 %
 
 \section{Introduction}
 Meteoritic stellar grains are solid samples of stars that can be studied in
 terrestrial laboratories. Condensed in the cooling gas outflows from ancient
 stars, they became part of the interstellar medium from which the Solar
 System formed some 4.6 billion years ago.
 Because they were encapsuled in primitive meteorites
 they remained ideally uncontaminated by Solar System material.
 The highly unusual isotopic ratios, with respect to
 solar, found in these ``stardust'' grains indicate that they are of stellar
 origin and can therefore be used as a diagnostic tool for verifying
 predictions from models of stellar evolution and nucleosynthesis
 \citep[cf.][]{anders93,zinner98,clayton04,lugaro:05}. Stardust
 grains come in many mineralogical flavours but of particular interest in this work
 are silicon carbide (SiC) grains.
 \newline \indent
 ``Mainstream'' SiC grains ($>$90$\%$ of stellar SiC grains) contain isotopic
 abundances of heavy elements characteristic
 of the {\it slow} neutron capture process (the {\it s }process) and are thus
 believed to have originated from asymptotic giant
 branch (AGB) stars \citep{lugaro:99,lugaro:03a}, which show enrichments at their
 surface of
 {\it s-}process elements such as Zr, Ba, and even the unstable Tc
 \citep{merrill52,smith90,busso01}.
  These stars are evolved giants in the final nuclear burning stage of
 evolution \citep[see ][ for a recent review]{herwig05}. Briefly, the AGB
 phase is important because of the occurrence of instabilities of the He-burning
 shell, known as thermal pulses (TPs). Inbetween TPs the H-shell provides most of the
 stellar luminosity. After the occurrence of a thermal pulse the third
 dredge-up (TDU) may occur, where the products of the partial He-burning such as
 $^{12}$C, along with other nucleosynthetic products such as {\it s-}process elements, are
 mixed to the stellar surface.
 \newline \indent
 An interesting
 nucleosynthetic product of AGB stars is the radioactive isotope $^{26}$Al
 (T$_{1/2}$ = 0.717 Myr). $^{26}$Al is produced by proton captures on
 $^{25}$Mg during H burning, when it can also be
 consumed by proton captures, depending on the rate of the $^{26}$Al+$p$ reaction,
 and destroyed
 by neutron captures in the thermal pulse because of the relatively high
 $^{26}$Al(n,p)$^{26}$Mg and
 $^{26}$Al(n,$\alpha$)$^{23}$Na rates. Neutrons in the TP are provided by the
 $^{22}$Ne($\alpha$,n)$^{25}$Mg reaction if the temperature exceeds $\approx$ 300 million
 K. When the TDU occurs the $^{26}$Al in the thin top layer of the intershell
 region not involved in the convective pulse, together with the $^{26}$Al
 that survived neutron captures in the TP, is carried to the surface. For a
 detailed description of the nucleosynthesis of $^{26}$Al in AGB stars see
 \citet{mowlavi00}.
 The combination of nucleosynthesis in the intershell and the occurrence of
 the TDU allows
 AGB stars of masses roughly between 1.5 \msun\ and 4 \msun\ to eventually become
 carbon rich \citep{groenewegen:95}, which is a
 necessary condition for the formation of SiC grains. Al is incorporated
 in SiC grains as a trace element.
 
 \begin{figure}
    \centering
    \includegraphics[angle=-90,width=9cm]{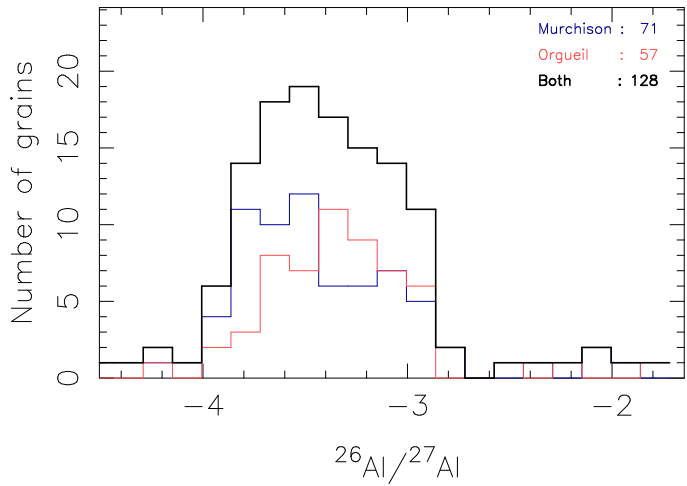}
       \caption{Histogram of binned $^{26}$Al/$^{27}$Al ratios measured in
                mainstream SiC stardust grains found in the Murchison and
                Orgueil meteorites.
 	       The total number of grains
                for which the $^{26}$Al/$^{27}$Al ratios were measured is
                indicated in the upper right hand corner.
               }
        \label{Figgraindistr}
 \end{figure}
 
 Since the abundance of Mg in SiC grains is much lower than that of Al,
 excesses in $^{26}$Mg, together with solar $^{25}$Mg/$^{24}$Mg ratios,
 observed in meteoritic stardust SiC grains are a measure of the
 $^{26}$Al abundance at the time and place of formation of the grain.
  Note that if spallation reactions on the grains during their residence
 time in the interstellar medium had been responsible for the observed huge excesses in
 $^{26}$Mg, this would also result in variations of the $^{25}$Mg/$^{24}$Mg
 ratios, which are not observed.
 The distribution of the available $^{26}$Al/$^{27}$Al data for
 mainstream SiC grains \citep{hoppe94,huss97}
 is presented in Fig. \ref{Figgraindistr}. The majority of the grains are distributed between
 ratios of $\approx$ 10$^{-4}$ to 2 $\times$ 10$^{-3}$.
 The $^{26}$Al/$^{27}$Al ratio has also recently been measured in a few SiC grains of
 the rare type Z, which are believed to have originated in low-Z AGB stars
 \citep{zinner07}. The data points cover the same spread as the
 mainstream grains and low metallicity AGB models give similar results to our $\approx$ solar
 metallicity models. Hence, our discussion can be applied to Z grains as well.
 
 The aim of this paper is to analyse in detail the effect of the uncertainty
 in the $^{25}$Mg($p,\gamma$)$^{26}$Al and $^{26}$Al$_g$($p,\gamma$)$^{27}$Si
 reaction rates on the production of $^{26}$Al in carbon rich AGB stars and
 compare the model predictions to the data obtained from mainstream SiC grains. While
 the $^{25}$Mg$+p$ is uncertain by a factor of $\approx$ 2 at the temperature of interest
 for H burning during the AGB stage ($\approx$ 60 million K), the $^{26}$Al$+p$
 reaction rate is uncertain by four orders of magnitude (see Fig.
 \ref{Figrates})
 due to possible contributions from as yet unobserved low-energy resonances
 \citep{iliadis01,angulo:99}. We evaluate the effect of such
 huge
 uncertainties on the outputs of AGB models and see if any constraints can be
 derived from comparison of these models to the SiC data. Stellar model
 uncertainties will also be briefly discussed.
 \newline \indent
 The study of radioactive isotopes in stardust grains is of interest also
 because they represent clocks to measure the timescale for the formation of
 dust around AGB stars \citep{zinner06a,davis06}. It is still very difficult
 to pin down the exact
 mechanism by which dust is formed around AGB stars, see for example
 discussion in \citet{nuth06}, and any insight on the
 timescale of grain formation therefore represents a
 useful constraint.
 \newline \indent
 We also note that $^{26}$Al is produced by proton captures at the base
 of the envelope (hot bottom burning, HBB) in intermediate mass AGB stars ({\it M}
 $\ga$ 5 {\it M}$_{\sun}$) and in AGB stars of lower masses if extra
 mixing is invoked, which is usually done to explain the
 composition of a particular fraction of stardust oxide grains
 showing a strong depletion of $^{18}$O along with high $^{26}$Al \citep{nittler97,nollett03}.
 Here we present
 only one model of an intermediate mass AGB star because
 they are not main producers of SiC stardust. In fact, HBB prevents massive,
  metal-rich (Z $>$ 0.004) AGB stars
 from developing a carbon rich envelope by converting C into N. Moreover,
 the isotopic signatures of C, N, Si, and heavy elements in mainstream SiC
 grains cannot be reconciled with massive AGB parent stars
 \citep{lugaro:99,lugaro:03a}. For example, HBB produces $^{12}$C/$^{13}$C
 in the range 3 to 10, while the mainstream SiC range is 20 to 100 and
 high neutron densities in the thermal pulses produce enhancements in the
 neutron-rich isotope $^{96}$Zr, which is instead observed to be depleted
 in mainstream SiC grains. We have not
 included extra mixing in our models, but
 will discuss its possible implications in \S 4, based on the models of
 \citet{nollett03}.
 In \S 2 we present the details of the methods and models we have used.
 In \S 3 we show the results we obtained, which we discuss in \S 4.
 We conclude with \S 5 where we present a summary and our main conclusions.
 
 
 \section{Methods and Models}
 \subsection{Evolutionary and Nucleosynthesis codes}
 We calculate the nucleosynthesis with a detailed post-processing code for
 which the stellar structure inputs were calculated beforehand.
  The stellar structure program we use has a small network of
 6 species: H, $^{3}$He, $^{4}$He, C, N, and O \citep[see e.g.
 ][]{lattanzio86}, involved in the
 main energy generation. To compute abundances of more
 species we use a post-processing code. The code
 is described in detail in \citet{lattanzio96}. Briefly:
 we input the structure (e.g. temperature, density, details of the convective
 regions, position of the H and He-burning shells as a function of interior
 mass and time) from the stellar evolution code to compute the
 abundances of species not involved in energy generating reactions.
 The post-processing step computes its own mass mesh with sufficient
 resolution in each burning shell (around 25 mass shells) to adequately
 resolve changes in abundances owing to nuclear burning. Because we
 assume the structure is fixed in the post-processing step we assume
 that the extra species and reaction rates added do not change the
 structure. For this reason we concentrate on uncertainties for
 reactions that produce negligible energy, such as those involved
 in the NeNa and MgAl chains, and neutron capture reactions.
 The details of the nucleosynthesis network are outlined in \citet{lugaro04},
 however, we remind the reader that we include 59 light nuclei and 14
 iron-group species. We also add the fictional particle $g$ to count the
 number of neutron captures occurring beyond $^{62}$Ni
 \citep{lattanzio96,lugaro04} in a similar manner to \citet{jorissen89}.
 
 The bulk of our 527 reaction rates are from the REACLIB tables
 \citep{thielemann86} updated as described in \citet{lugaro04} and
 \citet{karakas06}.
 \newline \indent
 In our calculations we used models of 1.75 {\it M}$_{\sun}$ with a
 metallicity ({\it Z}) of 0.008, of 3 {\it M}$_{\sun}$ with {\it Z} = 0.02,
 0.012, and 0.008 and of 5 {\it M}$_{\sun}$ with {\it Z} = 0.02 and the mass
 loss prescription from \citet{vassiliadis93}. The 1.75 {\it
 M}$_{\sun}$ and 3 {\it M}$_{\sun}$ models produce a carbon over oxygen ratio in excess of unity.
 More information on these stellar models can be found in
 \citet{lugaro:03b}, \citet{karakas06}, and \citet{karakas:07}.
 Models of low-mass AGB
 stars of approximately solar metallicity are shown to be the
 best to reproduce various features of stardust mainstream SiC grains:
 from the He and Ne composition \citep{gallino90}, to the
 $^{12}$C/$^{13}$C ratios, which are the same as observed in Galactic C stars
 with metallicity close to solar
 \citep{hoppe97}, to the heavy element compositions \citep{lugaro:03a}.
 Consequently, we focus on our AGB models of low mass and {\it Z} close to solar
 that become carbon rich. The initial abundances we
 use were taken from \citet{anders89}
 for {\it Z} = 0.02 and we assume scaled solar for the {\it Z} = 0.008
 models. The {\it Z} = 0.012 models are based on the \citet{asplund05} and
 \citet{lodders03} solar
 abundance table which
 prescribes a solar metallicity of {\it Z} = 0.012.
 
 \subsection{The reaction rates}
\begin{figure}
    \centering
    \includegraphics[angle=-90,width=9cm]{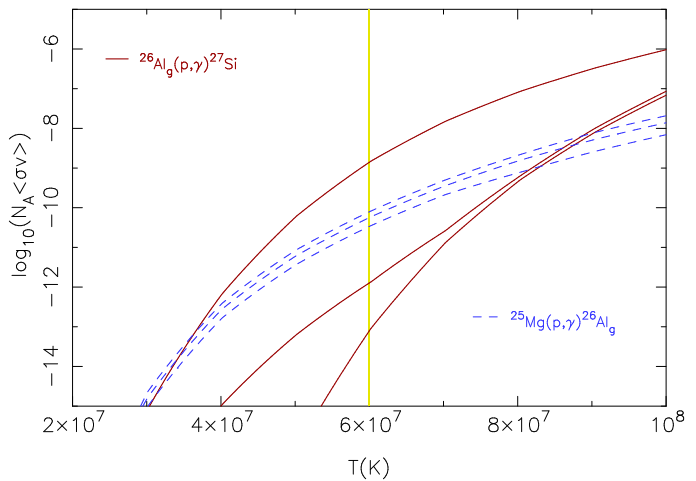}
       \caption{The lower limit (bottom line), recommended value
                (middle line), and upper limit (upper line) of the
                 $^{25}$Mg($p,\gamma$)$^{26}$Al$_g$ (dashed lines) and
                 $^{26}$Al$_g$($p,\gamma$)$^{27}$Si (solid lines) reaction rates.
 		The vertical line
                 denotes the approximate temperature of interest
 		(T $\approx$ 6$\times$10$^{7}$ K).}
          \label{Figrates}
    \end{figure}

 The rates of interest are those of the $^{25}$Mg($p,\gamma$)$^{26}$Al$_g$,
 $^{25}$Mg($p,\gamma$)$^{26}$Al$_m$, and $^{26}$Al$_g$($p,\gamma$)$^{27}$Al
 reactions.
 $^{26}$Al$_g$ is the ground state of $^{26}$Al, whereas
  $^{26}$Al$_m$ is the metastable state with a half life of 6.3452 s.
 Since the metastable state of $^{26}$Al is very short lived, the
 $^{25}$Mg($p,\gamma$)$^{26}$Al$_m$ rate
 results in the production of
 essentially no $^{26}$Al, but rather $^{26}$Mg.  Whenever we use
 $^{26}$Al with no subscript we mean
 the total sum of the metastable and ground state of $^{26}$Al. In
 practice this means $^{26}$Al$_g$, because the isomeric state is
 very unstable.
 The rates were taken from \citet{iliadis01}.
 We calculated models using all the nine combinations of the lower limits (LL),
 recommended values (RC),
 and upper limits (UL) of the rates,  where the rate errors of the
 reactions $^{25}$Mg($p,\gamma$)$^{26}$Al$_g$ and
 $^{25}$Mg($p,\gamma$)$^{26}$Al$_m$ are correlated. This assumption is
 justified since the uncertainties arise from the entrance channel partial
 width which is common to both interactions.
 The bulk of the rates
 in our code are calculated using
 fits in the 7-coefficient format of REACLIB; the rates of
 interest were read directly from the tabulated rates.
 \newline \indent
 Figure \ref{Figrates} shows the LL, RC, and UL rates for the
 $^{25}$Mg($p,\gamma$)$^{26}$Al$_g$ and $^{26}$Al$_g$($p,\gamma$)$^{27}$Si
 reactions for the temperature range dominant in the H burning
 shell during the AGB phase, where 6$\times$10$^{7}$ K
 is approximately the peak temperature for all of the used models, except the
 one with 5 {\it M}$_{\sun}$ and {\it Z} = 0.02 where it is 8$\times$10$^{7}$ K.
 The upper limit of the $^{26}$Al$_g$($p,\gamma$)$^{27}$Si rate is up to four
 orders of magnitude higher than the lower limit and results, as shown below,
 in a two orders of magnitude lower abundance of $^{26}$Al.
  These large uncertainties
 arise from as yet undetected low-energy resonances. In particular, an expected
 resonance at a center-of-mass energy of 94 keV, with a predicted upper limit of
 its strength of $\omega\gamma\!<$10$^{-8}$ eV, seems to play the most
 important role.
 
 \section{Results}
 
    \begin{figure}
    \centering
    \includegraphics[angle=-90,width=9cm]{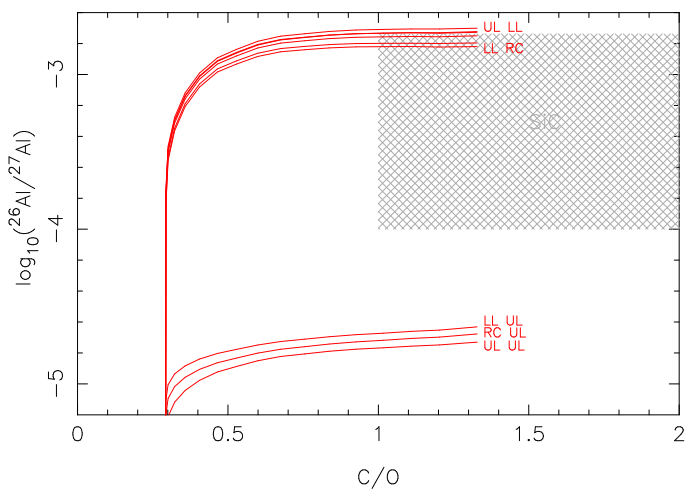}
       \caption{Surface $^{26}$Al/$^{27}$Al ratio versus C/O ratios for a star of mass 3
                {\it M}$_{\sun}$, metallicity {\it Z} =
                0.02, and a partial mixing zone (PMZ) of 1$\times$10$^{-3}$
                (see \S \ref{otheruncertainties}) using all combinations of
                upper, lower and recommended
                values for the rates under consideration. At the end of
                some selected lines a label indicates the values used for the
 	       $^{25}$Mg($p,\gamma$)$^{26}$Al$_g$ and
                $^{26}$Al$_g$($p,\gamma$)$^{27}$Si rates, in that order, where LL = lower
                limit, RC = recommended, and UL = upper limit.
 	       The grey crosshatched box is a schematic representation of
                the range of the $^{26}$Al/$^{27}$Al values observed in
                stardust maintream
                SiC grains (see Fig. \ref{Figgraindistr}), which can form
		only when C/O$>$1.}
          \label{Figallyield}
    \end{figure}
 
Figure \ref{Figallyield} shows our predictions for the $^{26}$Al/$^{27}$Al
ratio at the surface of the star as a function of C/O for the 3 {\it
M}$_{\sun}$ $Z$=0.02 model and all nine different combinations of the LL,
RC, and UL of the rates. All the lines show a very similar trend: the
$^{26}$Al/$^{27}$Al ratio increases sharply with the first few TDUs and
then becomes approximately constant. This is because during the first few
TPs the mass fraction of $^{26}$Al dredged up to the surface is
6.2$\times$10$^{-5}$ in the small region ($\approx$ 10$^{-4}$ {\it
M}$_{\sun}$) of the H-burning ashes \citep[region A of ][]{mowlavi00} and
$\approx$ 1.8$\times$10$^{-5}$ in the rest of the intershell \citep[region
D of ][]{mowlavi00}. \citep[See also Table 2 of ][]{lugaro01}. However, as
the pulse number increases, the temperature at the base of the convective
intershell region increases, and the $^{22}$Ne($\alpha$,n)$^{25}$Mg
reaction becomes more active, this frees up more neutrons resulting in more
$^{26}$Al destruction by neutron capture. The abundance of $^{26}$Al in the
last computed TP is in fact $\approx$ 2$\times$10$^{-8}$ in the intershell.
The TDU of this small amount of $^{26}$Al is just enough to ensure that the
small fraction ($\approx$ 5\%) of $^{26}$Al at the stellar surface that
decays during the interpulse period is replenished. As a result the
prediction lines flatten. 
 
Two main classes of prediction lines can be distinguished:
 
\begin{enumerate}
\item{When considering the models computed using the RC and LL values
for the $^{26}$Al$_g$($p,\gamma$)$^{27}$Si rate the variation among these
is of a factor of a mere 1.32. Of this, a factor of $\approx$ 1.26 is
derived when varying the $^{25}$Mg($p,\gamma$)$^{26}$Al$_g$ reaction
rate between the LL and the UL, while a factor of $\approx$ 1.05 is
derived when varying the $^{26}$Al$_g$($p,\gamma$)$^{27}$Si reaction
rate between the LL and the RC rate. All the lines lie around the upper
end of the SiC grain data range.}       
\item{When considering the models computed using the UL values for the
$^{26}$Al$_g$($p,\gamma$)$^{27}$Si rate, the $^{26}$Al/$^{27}$Al ratio is
roughly two orders of magnitude smaller than the ratios computed using the
RC and the LL values of the rate, hence falling below the range of values
observed in mainstream SiC grains.}
\end{enumerate}

 \begin{figure}
    \centering
    \includegraphics[angle=-90,width=9cm]{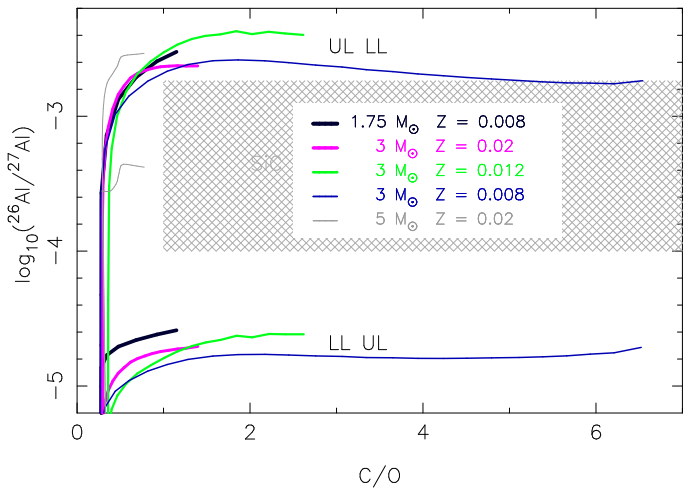}
    \caption{$^{26}$Al/$^{27}$Al abundance ratios for models of
                different mass and metallicity. Two calculations are shown for each
                model: the
                top line represents the reaction rate
 	       combination: UL LL and the bottom line the reaction rate
                combination: LL UL (see Fig. \ref{Figallyield}). The fact that the 3 {\it
                  M}$_{\sun}$ $Z$=0.02 line is slighlty higher than shown in
                Fig. \ref{Figallyield} is due to the fact that a partial mixing zone was included
                 in the calculations shown in Fig. \ref{Figallyield}, while no partial mixing zone
                is included in the calculations shown in this figure.}
          \label{FigallMnZnopmz}
    \end{figure}
  
 In Fig. \ref{FigallMnZnopmz} we present the results for different masses
 and metallicities. We can see the same behaviour here as described above:
 the upper limit of the $^{26}$Al$_g$($p,\gamma$)$^{27}$Si reaction rate always
 results in a factor of $\approx$ 100 less $^{26}$Al, except for the 5 {\it
 M}$_{\sun}$ model, in which case the difference is of a factor of
 $\approx$ 10. Note that the production of $^{26}$Al in this massive AGB
 model is due to the operation of the second and third dredge-up and $not$
 due to hot bottom burning because the temperature at the bottom of the
 convective envelope is not high enough to produce $^{26}$Al
 \citep[see discussion in][]{lugaro:07}. This explains why our detailed
 model shows a large effect in the $^{26}$Al abundance as a result of the
 uncertainties in the $^{26}$Al$_g$($p,\gamma$)$^{27}$Si reaction rate, while
 the synthetic models of \citet{izzard07}, which only describe the effect
 of hot bottom burning, do not show this effect.

 All the models computed using the RC or LL for the
 $^{26}$Al$_g$($p,\gamma$)$^{27}$Si reaction rate produce
 $^{26}$Al/$^{27}$Al ratios at the upper end of the observed distribution.
 Similar results were
 obtained (for the RC case) by \citet{mowlavi00} and \citet{zinner07} for several models of
 different masses and {\it Z}, and by \citet{cristallo06} for a 2 {\it
 M}$_{\sun}$ {\it Z} = 0.015 model.
 
 \subsection{Other uncertainties}
 \label{otheruncertainties}
 
 Some uncertainty is introduced by the neutron capture reaction rates. We use
 for these rates the estimates of \citet{koehler:97} and
 \citet{skelton:87}. These works suggest for the $^{26}$Al$_g$($n,p$)$^{26}$Mg rate a value similar to
 that of \citet{caughlan:88}, $\approx$ 250 mbarn (at 23 keV, taken as typical temperature in the
 thermal pulse) and for the $^{26}$Al$_g$($n,\alpha$)$^{23}$Na rate a value roughly a factor of two higher
 than \citet{caughlan:88}, $\approx$ 180 mbarn (at 23 keV). The
 $^{26}$Al$_g$($n,\gamma$)$^{27}$Al rate is much smaller than the
 aforementioned neutron capture channels: $\approx$ 4.5 mbarn \citep[at 23 keV, ][]{bao:00}. We
 conservatively estimated
 the uncertainties of these rates to be of a factor of two both above and below the
 values we use. When we multiplied the rate of all
 $n$-capture rates on $^{26}$Al by a factor of two we ended up with half as much $^{26}$Al,
 similarly when we decreased the $n$-capture rates by a factor of
 two, we ended up with twice as much $^{26}$Al. Overall, the uncertainty is at most of a factor
 of four.
 \newline \indent
  Since neutrons released by the $^{22}$Ne($\alpha$,n)$^{25}$Mg reaction
 destroy $^{26}$Al, the uncertainty of this reaction rate
 introduces another uncertainty in the $^{26}$Al abundance.
 Between the lower and upper limit
 \citep[a range of about a factor of 15, ][]{karakas06}
 of this rate the
 $^{26}$Al abundance varies with a factor of 1.8 for the 3 {\it
 M}$_{\sun}$ and {\it Z} = 0.02 model.
 \newline \indent
  Note that the uncertainties of a
 factor of $\approx$ 20 \citep{iliadis01} in the $^{27}$Al$+p$ reaction rate do
 not change the $^{27}$Al abundance because this reaction is too slow in the
 H-burning shell of AGB stars.
 \newline \indent
 As for stellar uncertainties:
 we artificially included a
 partial mixing zone in the top layers of the intershell in the way described in detail by
 \citet{lugaro04}. There, protons
 combine with $^{12}$C to make $^{13}$C, which releases neutrons for the {\it s }process via
 $^{13}$C($\alpha$,n)$^{16}$O during the interpulse periods, however, the
 nature of the mixing is still debated \citep{herwig05}.
 Varying the size of the $^{13}$C pocket from 0 to 2$\times$10$^{-3}$
 introduces a spread in the $^{26}$Al/$^{27}$Al ratio of at most a factor of 1.4.
  Varying the proton profile in the
 partial mixing zone changes the relative importance of the
 $^{14}$N-poor and $^{14}$N-rich regions of the pocket. As shown in Fig. 1
 of \citet{goriely:00}, in the $^{14}$N-poor region $^{26}$Al is completely
 destroyed, while in the $^{14}$N-rich region it is destroyed to an
 abundance of the order of 10$^{-7}$ in number. Thus, changes in the
 proton profile would not have a significant impact on the overall
 $^{26}$Al abundance in these stars.
 \newline \indent
 Mass loss and third dredge-up are very uncertain physical features of AGB
 stars. However, we expect that changes in the TDU efficiency, caused by
 variations in the input physics, and changes of the mass loss values will
 not affect our results significantly. This is because we are looking at
 C-rich stars and the C/O$>$1 constraint sets the dilution factor in our
 models to $\approx$ 1 part of intershell material to $\approx$ 30 parts of
 envelope material. This dilution factor would necessarily produce
 $^{26}$Al/$^{27}$Al $\approx$ 10$^{-3}$, indepently of which TDU efficiency
 and mass loss rate have been employed to achieve it.
 
  Extra-mixing processes due to rotation or other mechanisms may also
 be at work in these stars. Extra mixing is the hypothesis that some of
 the material from the convective envelope is mixed into the radiative
 layer
 that resides on top of the hydrogen shell. It is also commonly referred
 to as: ``deep mixing'' and ``cool bottom processing''. Extra mixing was
 originally introduced into the first giant phase of evolution to explain
 several abundance peculiarities, including lower than predicted
 $^{12}$C/$^{13}$C ratios in first giant branch stars \citep[see
 e.g.][]{sweigart:79}. Extra mixing has been invoked for AGB stars to
 explain the $^{12}$C/$^{13}$C ratio in C-rich stars \citep[][]{abia00},
 as well as in mainstream and Z SiC grains \citep[][]{zinner06b}, the O
 composition of AGB stars \citep[see e.g][]{wasserburg:95} and the O and
 Al composition of a fraction of stellar oxide grains
 \citep[][]{nollett03}. The possible effect of extra-mixing on the
 $^{26}$Al abundance will be discussed in \S4.2.

 
 \section{Discussion}
 
 With the exception of the 5 {\it M}$_{\sun}$ model, the models presented in Fig. \ref{FigallMnZnopmz}
 should represent the set of stars, in terms of
 mass and metallicity, responsible for mainstream SiC stardust grains.
 However, the spread of the $^{26}$Al/$^{27}$Al SiC grain data is much larger than that of
 the models. The spread of the data is not an error range, but arises from considering the large
 number of single grains (128, to be exact, see Fig. \ref{Figgraindistr}).
 The $^{26}$Al/$^{27}$Al abundance ratios range a factor of $\approx$ 20 in
 the data, whereas the models show at best a spread of about a factor of 2.
 We propose two scenarios to explain the observed range:
 
 \subsection{The $^{26}$Al$_g$$+p$ reaction rate corresponds to its current LL or RC
 value}
 
 In this case we need to explain the fact that the data extends below the
 prediction lines, which may reflect the fact that stardust SiC grains have
 a long interval of formation time between a time close to when TDU has
 enriched the stellar winds with $^{26}$Al and approximately two million
 years later. This time delay would allow the decay of $^{26}$Al into
 $^{26}$Mg, the latter of which is not incorporated in the grains. Since
 the mechanism by which the large (up to 25 $\mu$m) stardust SiC grains
 found in meteorites are formed around single AGB stars is not understood,
 and in any case it probably does not involve timescales longer than $\sim
 10^5$ yr \citep{nuth06}, we speculate upon different possibilities.
 \newline \indent
 A long interval of formation time could be achieved if the grains were
 formed either (i) in the winds of extrinsic carbon stars formed in a
 binary system (in this case the absence of $^{26}$Al would be conceptually
 equivalent to that of Tc in these stars) or, perhaps, (ii) in a
 long-lived circumbinary disk that formed after one of the binary stars
 evolved through the AGB phase.
 Note that circumbinary disks
 have already been proposed as the site of origin of at least some meteoritic stardust in order to
 explain the large observed sizes of the grains $>$ 0.5 $\mu$m \citep{jura97}.
 The long timescale of stardust grain formation proposed here
  is not ruled out by the analysis of radioactive heavy
 nuclei in mainstream SiC, which have been discussed in detail by
 \citet{davis06}.
 
 \subsection{The $^{26}$Al$_g$($p,\gamma$)$^{27}$Si reaction rate
 corresponds to its current UL}
 
 In this case we need to explain the fact that the data extends above the prediction lines, which
 may reflect the occurrence of extra-mixing processes.
 Extra mixing could produce $^{26}$Al/$^{27}$Al
 ratios higher than our predictions and the spread in the
 data could be achieved by a range of temperatures (i.e. depths) to which the mixed material is
 exposed. The main problem with invoking extra mixing is that we still miss the physical mechanism
 behind the process,  even if some steps forward in this search have been made recently
 \citep{eggleton:06,charbonnel:07,busso07}.
 
  Note that assuming the UL for the
 $^{26}$Al$_g$($p,\gamma$)$^{27}$Si reaction rate does not affect $^{26}$Al production in supernovae,
 since most of this nucleus is produced during the explosion where the main destruction channel
 for $^{26}$Al is neutron capture \citep[][]{limongi06}. Thus, the present scenario would not be
 in contrast with the match to the observed live $^{26}$Al in the Galaxy
 \citep[][]{diehl06}. On the other hand, production of $^{26}$Al in
 Wolf-Rayet stars \citep[][]{arnould06} and in
 intermediate-mass AGB stars by hot bottom burning would be affected \citep[for a detailed
 analysis on the latter see][]{izzard07}. As a consequence, the discussion on the origin of
 stardust spinel grain OC2 from a massive AGB star \citep{lugaro:07} would have to be revised.
 
 
 \section{Summary and conclusions}
 
 We have shown that the large uncertainty of the $^{26}$Al$_g$$+p$ reaction
 rate has important implications when comparing AGB models to the
 $^{26}$Al/$^{27}$Al ratios observed in stardust mainstream SiC grains. We
 have presented two scenarios to explain the observed distribution: one
 involves using the LL or the RC rate and invoking a relatively long timescale
 of grain formation, possibly connected to processes occurring in binary
 systems. The other involves using the UL of the rate and invoking
 extra-mixing processes.
 
 \begin{figure}
    \centering
    \includegraphics[angle=0,width=9cm]{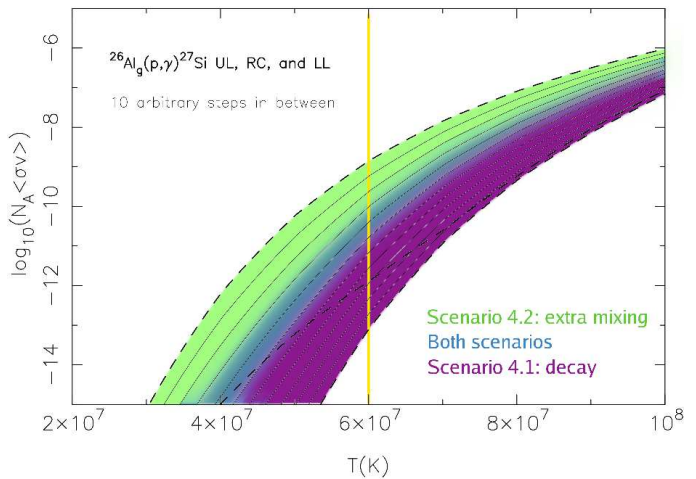}
    \caption{ Upper limit, recommended value, and lower limit of the $^{26}$Al$_g$($p,\gamma$)$^{27}$Si
              reaction rate (dashed lines) as in Fig. \ref{Figrates}, with 10
              logarithmically equidistant steps in between the LL and the UL (solid lines).
 	     The shaded areas represent roughly
 	     which value of the $^{26}$Al$_g$($p,\gamma$)$^{27}$Si reaction
 	     rate corresponds to which scenario. The vertical yellow line
              denotes the approximate temperature of interest (as in
	      Fig.\ref{Figrates}).}
          \label{Figscenarios}
 \end{figure}
 
 To check in details which scenario would correspond to which values of the
 rate we have computed the $^{26}$Al/$^{27}$Al ratios in the 3 \msun,
 $Z=$0.02 model for ten different rates calculated as logarithmically equidistant steps in
 between the LL and UL. The results are shown in Fig. \ref{Figscenarios}.
 The decay scenario has to be invoked with rate values ranging from the LL
 up to the 6$^{th}$ intermediate line from below.
 For rates from roughly the 8$^{th}$ intermediate line up to the UL
 the $^{26}$Al/$^{27}$Al ratio lies below the SiC data, and hence
 would imply the extra mixing scenario.
 
 Our work calls for more laboratory measurements of the $^{26}$Al/$^{27}$Al
 ratios in stardust SiC grains of different sizes and for improved measurements of the
 $^{26}$Al$_g$+$p$ reaction rates. A direct measurement  of the 94 keV resonance
 may prove challenging, considering the difficult preparation of a radioactive
 $^{26}$Al beam-stop target that will not degrade under intense proton bombardment.
 An interesting attempt at studying the nuclear structure of the corresponding compound
 level in $^{27}$Si by using the $^{26}$Al$_g$($^{3}$He,$d$)$^{27}$Si reaction has been reported
 in \citet{vogelaar:96}. Unfortunately, the contamination of their evaporated transmission
 target precluded the observation of transfers to any threshold states. A remeasurement of
 the transfer reaction by using an isotopically and elementally pure implanted $^{26}$Al
 target seems more promising.
 \newline \indent
 Once the estimate of the $^{26}$Al$_g$($p$,$\gamma$) reaction is less uncertain,
 we will be able to readdress the issue of $^{26}$Al production in AGB stars
 and deliver stronger conclusions on the implications for dust formation
 around AGB stars and extra-mixing processes.
  
 \begin{acknowledgements}
 ML gratefully acknowledges the support of NWO through the VENI grant. We thank Peter Hoppe and
 Gary Huss for providing us with the datatables of the grain measurements and
 for helpful and instructive discussion. We thank Alessandro Chieffi and Marco Limongi
 for discussion. AIK gratefully acknowledges the financial support of the
 Australian Research Council from the Discovery Project funding scheme
 (project number DP0664105).  We thank the referee, Marcel Arnould, for a
 detailed report which helped to improve the paper.
 \end{acknowledgements}
 

 \bibliographystyle{aa} 
 
 \end{document}